\documentclass[journal,twocolumn]{IEEEtran}
\IEEEoverridecommandlockouts
\usepackage{cite}
\usepackage{amsmath,amssymb,amsfonts}
\usepackage{hyperref}
\hypersetup{
    colorlinks,
    linkcolor={black},
    urlcolor={blue!80!black}
}
\usepackage{algorithm,algcompatible}
\DeclareMathOperator*{\argmin}{argmin}
\algnewcommand\INPUT{\item[\textbf{Input:}]}%
\algnewcommand\OUTPUT{\item[\textbf{Output:}]}%

\usepackage{graphicx}
\usepackage{textcomp}
\usepackage{xcolor}
\def\BibTeX{{\rm B\kern-.05em{\sc i\kern-.025em b}\kern-.08em
    T\kern-.1667em\lower.7ex\hbox{E}\kern-.125emX}}
    
\makeatletter
\newcommand{\setword}[2]{%
  \phantomsection
  #1\def\@currentlabel{\unexpanded{#1}}\label{#2}%
}
\makeatother
\usepackage{tikz}
\usepackage{xcolor}
\usepackage{pgfplots}
\usetikzlibrary{spy}
\pgfplotsset{compat=1.18} 
\renewcommand{\Bbb}{\mathbb}
\begin{document}

\title{Self-supervised Deep Hyperspectral Inpainting with the Sparsity and Low-Rank Considerations\\
}
\author{Shuo Li, Mehrdad Yaghoobi
\thanks{S. Li, and M. Yaghoobi are with the Institute for Digital Communications, School of Engineering, University of Edinburgh, EH9 3JE UK (e-mail: S.Li-91@sms.ed.ac.uk; m.yaghoobi-vaighan@ed.ac.uk).}
}

\maketitle

\begin{abstract}
Hyperspectral images are typically composed of hundreds of narrow and contiguous spectral bands, each containing information about the material composition of the imaged scene. However, these images can be affected by various sources of noise, distortions, or data losses, which can significantly degrade their quality and usefulness. To address these problems, we introduce two novel self-supervised Hyperspectral Images (HSI) inpainting algorithms: Low Rank and Sparsity Constraint Plug-and-Play (LRS-PnP), and its extension LRS-PnP-DIP, which features the strong learning capability, but is still free of external training data. We conduct the stability analysis under some mild assumptions which guarantees the algorithm to converge. It is specifically very helpful for the practical applications. Extensive experiments demonstrate that the proposed solution is able to produce visually and qualitatively superior inpainting results, achieving state-of-the-art performance. The code for reproducing the results is available at \url{https://github.com/shuoli0708/LRS-PnP-DIP}.
\end{abstract}


\section{Introduction}
Hyperspectral remote sensing has been widely applied to numerous applications such as astronomy, agriculture, environmental monitoring and earth observation. The hyperspectral images (HSI) are often captured by the satellite or airborne sensors, each presenting the samples at different time slots, using push-broom strategies along the flying pathway \cite{ortega2019hyperspectral}. The nature of the HSI acquisition system makes HSI a high resolution 3D data cube that covers hundreds or thousands of narrow spectral bands, conveying a wealth of spatio-spectral information. However, due to the influence of instrumental errors, imperfect navigation and atmospheric changes, practical HSI images can suffer from noise and missing pixels, or a lines of pixels. These issues can severely impact subsequent applications, making HSI inpainting a critical task in the field of remote sensing and earth observations.

\subsection{Hyperspectral Image Inpainting}
Hyperspectral image inpainting involves restoring missing or corrupted data in acquired HS images. In contrast to RGB images, inpainting of HSIs requires filling in a complex vector that contains extensive spectral information, rather than just a single pixel value. The additional complexity makes the already challenging task of HSI inpainting even more difficult. The primary objective of HSI inpainting is then to create visually convincing structures while ensuring that the texture of the missing regions is spectrally coherent.
HSI inpainting typically involves two main steps: a) estimating the missing or corrupted data using a model that utilizes information from the neighbouring spectral bands and spatially adjacent pixels, and b) refining the estimated data to ensure it is consistent with the overall spectral and spatial properties of the HS image.
The conventional approach \cite{zhuang2018fast} assumes that the spectral vectors of HSI always exist in some unknown low-dimensional subspaces, and the missing regions are estimated through projections. Nonetheless, the effectiveness of the technique is significantly diminished when a substantial number of pixels are missing, and it may even fail when the entire spectral bands are absent.
In the past decade, the success of deep learning has brought new opportunities on how to solve the HSI inpainting task. In \cite{wong2020hsi}, authors show that the missing pixels can be predicted by training deep neural networks. However, it has two drawbacks, it requires training the networks on some extensive datasets to achieve desired performance. 2) the inpainted areas tend to be over-smoothed, resulting in the loss of critical information such as sudden changes in the surface materials. Recently, researchers discovered that the network's structure itself may serve as a good inductive bias \cite{ulyanov2018deep} to regularise some image reconstruction tasks, \textit{e.g.} image inpainting. It is found that it is possible to learn the priors directly from the image, without the need for extensive training data . The proposed method, called Deep Image Prior (DIP), has been further developed and successfully applied to Hyperspectral (HS) images\cite{sidorov2019deep}, achieving state-of-the-art HS inpainting performance. Following \cite{sidorov2019deep}, the subsequent works \cite{2021_DIP_In_Loop,lai2022deep,wu2022adaptive} have shown that certain trained or untrained neural networks can be directly incorporated into iterative solvers to improve reconstruction accuracy, \textit{i.e.} as the regulariser. However, as it is pointed out in \cite{2021_DIP_In_Loop}, the conventional DIP method is susceptible to over-fitting, which requires either a suitable early stopping criterion or manual control of the learning rate. When DIP is utilized in a loop, this issue will be further exacerbated. The new HSI inpainting algorithm, which harvests the benefits of previous PnP methods with convergence guarantees, is sought.

\subsection{Sparsity and Low Rankness in HSI}
Although an HS has high dimensionality, it intrinsically exposes sparse and low-rank structures due to the high correlation between spectral channels and spatial pixels. This property has made sparse representation (SR) and low-rank (LR) based methods popular for the processing HSIs \cite{sparse_representation,low_rank}. SR relies on the key assumption that the spectral signatures of pixels approximately lie in a low-dimensional subspace spanned by representative pixels from the same class. With a given dictionary, SR allows for a sparsely decomposition of the HSIs into a linear combination of several atoms, thus exploiting the spectral similarity of HSIs. This assumption on intrinsic structures has been effective in various HSI tasks, including HSI classification \cite{SR-application_1}, denoising \cite{LR_SR_1}, and un-mixing\cite{SR-application_3}. It is worth mentioning that when the dictionary is not given, we can manually use the end-members of some pixels with pure spectra and build the dictionary, or learn the dictionary \cite{dictionary_learning} in an unsupervised/self-supervised way, using the sparsity as a regulariser for weights. 
The low rankness is another widely used regulariser in HSI processing. It assumes that pixels in the clean HSI image have high correlations in the spatial domain, thus capturing the spatial similarity and global structure of HSIs. The LR is often combined with SR to achieve better performance \cite{LR_SR_1,LR_SR_2}. For more detailed information about LR and SR, we refer readers to the review work in\cite{SR_LR_review}. In this paper, we utilise both of these two priors and finally propose two novel self-supervised deep learning-based HSI inpainting algorithms to predict the missing pixels from the noisy and incomplete observations. 

\subsection{Contribution}
The aim of this paper is to present an HS inpainting algorithm that leverages the powerful learning capability of deep networks, without requiring any external training data, \textit{i.e.} called self-supervised learning. Our approach offers several key contributions, including:
\begin{itemize}
\item We propose an approach to HSI inpainting that is able to address challenging scenarios where entire spectral bands are missing. 
\item A deep hyperspectral prior is utilized to enhance the inpainting performance, which achieves state-of-the-art performance by effectively leveraging the unique features of HSIs.
\item Under some mild assumptions, a fixed-point convergence proof is provided for the LRS-PnP-DIP algorithm (see Theorem 1). 
\item Thorough extended experiments, conducted using real data to validate the effectiveness of our proposed LRS-PnP and LRS-PnP-DIP algorithms, the superiority of proposed frameworks over existing traditional/learning-based methods are demonstrated.
\end{itemize}

The rest of this paper is organized as follows: section \textrm{II} provides an introduction to our approach, section \textrm{III} states our main convergence theory for the proposed LRS-PnP-DIP algorithm, followed by the implementation details in section \textrm{IV}. Section \textrm{V} discusses the experiment results.
Section \textrm{VI} finally concludes the paper with a brief discussion about potential future directions.

\section{Mathematical Formulations: LRS, LRS-PnP and LRS-PnP-DIP}
The task of HSI inpainting can be understood as reconstructing the clean image $X$ from a noisy and incomplete measurement $Y$, where an additive noise $N$ and masking operator $M$ are present:
\begin{equation} \label{eqn:problem setting}
 Y = M\{X\} +N
\end{equation}
The clean image $X \in\mathbb{R}^{q}$ (where $q =n_{r} \times n_{c} \times n_{b}$), with $n_{r}$ and $n_{c}$ representing the spatial dimensions of the image, and $n_{b}$ representing the total number of spectral bands. The operator $M\in \mathbb{R}^{q \times q}$ is a binary mask, where zero represents missing pixels and one represents observed and valid pixels. Thus, $M$ can be represented as a diagonal square matrix. $N$ is additive Gaussian noise of appropriate size. Usually, $M$ is provided, and the formulation \eqref{eqn:problem setting} represents a linear system that can be written as:
\begin{equation} \label{eqn:problem setting_2}
 \boldsymbol{y} = \rm M \boldsymbol{x} +\boldsymbol{n}
\end{equation}
In this equation, $\boldsymbol{x}$, $\boldsymbol{y}$, and $\boldsymbol{n}$ represent the vectorized forms of $X$, $Y$, and $N$, respectively, while $\rm M$ is a diagonal matrix.
We can obtain the recovered, \textit{i.e.} inpainted, image $\boldsymbol{x^*}$ by applying sparse representation to each image patch $P_i(\boldsymbol{x})$. To do so, we introduce an operator $P_i(\boldsymbol{x})$, which extracts the i-th patch from the image $\boldsymbol{x}$. Note that each image patch may cover only the valid pixels or may include the missing pixels, depending on the size of $P_i(\cdot)$. The inpainted image $\boldsymbol{x^*}$ can be obtained by solving the following optimization problem:
\begin{equation}
\begin{aligned}
 (\boldsymbol{x^*},\boldsymbol{\alpha^*}) = &\argmin_{\boldsymbol{\alpha},\boldsymbol{x}}   \gamma\Vert \boldsymbol{y} -\rm M\boldsymbol{x} \Vert_{2}^2 + w_{lr}\Vert \boldsymbol{x} \Vert_* + w_{s}\Vert \boldsymbol{\alpha} \Vert_1\\
&\textrm{s.t.} \quad \boldsymbol{x} = \Phi \boldsymbol{\alpha} 
\end{aligned}
\end{equation}
The objective in the expression above is composed of three terms, the first being the data fidelity term, which is weighted using the parameter $\gamma$. Since estimating $\boldsymbol{x}$ from $\boldsymbol{y}$ is inherently ill-posed, \textit{i.e.} more unknown variables than equations, the solution is non-unique. As a result, we introduce two additional "priors" to regularize the inpainting problem: low-rank and sparsity constraints. The second term penalise the solution $\boldsymbol{x}$ to be low-rank, which is typically employed as a surrogate for the rank minimization. The third term restricts the missing pixels to be generated from the subspace approximated by the valid pixels. Similarly, we weight these terms with parameters $w_{lr}$ and $w_{s}$, respectively. The sparse representation problem is solved using a given dictionary $\Phi$. Specifically, $\Phi$ can be constructed by either using the end-members of some pixels with pure spectra, or learning from the noisy pixels in the observations. It is here exclusively learned from the noisy pixels in the observations using online dictionary learning \cite{dictionary_learning}.
By adopting the augmented Lagrangian and introducing the auxiliary variable $\boldsymbol{u}$\cite{ADMM}, problem \eqref{eqn:problem setting_2} can be rewritten as:
\begin{equation} \label{opmization_problem}
\begin{aligned}
(\boldsymbol{x^*},\boldsymbol{\alpha^*}) = &\argmin_{\boldsymbol{\alpha},\boldsymbol{x}} \gamma \Vert \boldsymbol{y} -\rm M\boldsymbol{x} \Vert_{2}^2 + w_{lr}\Vert \boldsymbol{u} \Vert_*
 + w_{s}\sum_i \Vert \boldsymbol{\alpha}_i \Vert_1. \\
 &+\frac{\boldsymbol{\mu}_1}{2}\Vert \sum_i(P_i(\boldsymbol{x})-\Phi \boldsymbol{\alpha}_i) +\frac{\boldsymbol{\lambda}_1}{\boldsymbol{\mu}_1} \Vert_{2}^2\\
&\textrm{s.t.} \quad \boldsymbol{x} = \boldsymbol{u}
\end{aligned}
\end{equation}
The Lagrangian multiplier and penalty term are denoted by $\boldsymbol{\lambda}_1$ and $\boldsymbol{\mu}_1$, respectively. By using the alternating direction method of multipliers (ADMM), we can sequentially update the three variables $\boldsymbol{\alpha}$, $\boldsymbol{u}$, and $\boldsymbol{x}$ to solve problem \eqref{opmization_problem}: \\
1) \textit{Fixing $\boldsymbol{u}$ and $\boldsymbol{x}$, and updating $\boldsymbol{\alpha}$}:
\begin{equation} \label{alpha}
\begin{aligned}
 {\boldsymbol{\alpha}}^{k+1} = \argmin_{\boldsymbol{\alpha}} \frac{\boldsymbol{\mu}_1^k}{2} \sum_i\Vert(P_i(\boldsymbol{x}^k) + \frac{\boldsymbol{\lambda}_1^k}{\boldsymbol{\mu}_1^k})-\Phi \boldsymbol{\alpha}_i \Vert_{2}^2 \\+  w_{s}\sum_i \Vert \boldsymbol{\alpha}_i \Vert_1 \\
\end{aligned}
\end{equation}
The problem in equation \eqref{alpha} is a patch-based sparse coding problem which can be solved using iterative solvers. In our approach, we employ PnP-ISTA \cite{PnP-ISTA}, which has demonstrated superior performance compared to the conventional ISTA\cite{ISTA}. Let us denote the first term in equation \eqref{alpha} as $f = \frac{\boldsymbol{\mu}_1^k}{2} \sum_i\Vert(P_i(\boldsymbol{x}^k) + \frac{\boldsymbol{\lambda}_1^k}{\boldsymbol{\mu}_1^k})-\Phi \boldsymbol{\alpha}_i \Vert_2^2$. The entire process can then be replaced by an off-the-shelf denoiser $\mathcal{D}$, operating on the gradient of $f$\cite{PnP-ISTA}. Each iteration takes the following form: \\
\begin{equation}
\begin{aligned}
 {\boldsymbol{\alpha}^{k+1}} = \mathcal{D}(I-\nabla f)(\boldsymbol{\alpha}^{k})
\end{aligned}
\end{equation}
2) \textit{We can update $\boldsymbol{u}$ in this setting by}:
\begin{equation}
\begin{aligned}
  {\boldsymbol{u}^{k+1}} = \argmin_{\boldsymbol{u}} w_{lr}\Vert \boldsymbol{u} \Vert_* + \frac{\boldsymbol{\mu_2}^{k}}{2}\Vert (\boldsymbol{x}^k+\frac{\boldsymbol{\lambda_2}^k}{\boldsymbol{\mu}_2^k}) - \boldsymbol{u} \Vert_{2}^2
\end{aligned}
\end{equation}
The problem can be solved using the Singular Value Thresholding (SVT) algorithm \cite{cai2010singular}. Specifically, element-wise soft shrinkage is applied to the singular value of $(\boldsymbol{x}^k + \frac{\boldsymbol{\lambda}_2^k}{\boldsymbol{\mu}_2^k})$, as follows:
\begin{equation}
\begin{aligned}
  {\boldsymbol{u}^{k+1}} = SVT(\boldsymbol{x}^k+\frac{\boldsymbol{\lambda}_2^k}{\boldsymbol{\mu}_2^k})
\end{aligned}
\end{equation}
In the proposed LRS-PnP-DIP algorithm, this step is substituted with a deep image prior (DIP) $f_\theta(\boldsymbol{z})$, where $\theta$ denotes the network parameters that need to be updated, and the input $\boldsymbol{z}$ is set to $\boldsymbol{x}^k + \frac{\boldsymbol{\lambda}_2^k}{\boldsymbol{\mu}_2^k}$, i.e., the latent image from the previous iterations:
\begin{equation}
\begin{aligned}
 {\boldsymbol{u}^{k+1}} = f_\theta(\boldsymbol{x}^k + \frac{\boldsymbol{\lambda}_2^k}{\boldsymbol{\mu}_2^k})
\end{aligned}
\end{equation} \\
3) \textit{Fixing $\alpha$ and $\boldsymbol{u}$, and updating $\boldsymbol{x}$}: \\
\begin{equation}
\begin{aligned}
  {\boldsymbol{x}^{k+1}} = \argmin_{\boldsymbol{x}}  \gamma\Vert \boldsymbol{y} -\rm M\boldsymbol{x} \Vert_{2}^2 + \sum_i\Vert(P_i(\boldsymbol{x}) + \frac{\boldsymbol{\lambda}_1^k}{\boldsymbol{\mu}_1^k})\\
  -\Phi\boldsymbol{\alpha}_i^{k+1} \Vert_{2}^2 
  + \frac{\boldsymbol{\mu}_2^k}{2}\Vert (\boldsymbol{x}+\frac{\boldsymbol{\lambda}_2^k}{\boldsymbol{\mu}_2^k}) - \boldsymbol{u}^{k+1} \Vert_{2}^2 \\
\end{aligned}
\end{equation}
Closed-form solution for $\boldsymbol{x}$ exists as follows: \\
\begin{equation} \label{x_closed_form}
\begin{aligned}
  &{\boldsymbol{x}^{k+1}} = (\gamma \rm M^T\rm M + \boldsymbol{\mu}_1^k \sum_i P_i^TP_i        +\boldsymbol{\mu}_2^k\rm I)^{-1} \\ 
   &( \gamma \rm M^T\boldsymbol{y}+ \boldsymbol{\mu}_1^k \sum_i P_i\Phi \boldsymbol{\alpha}_i^{k+1} + \boldsymbol{\mu}_2^k \boldsymbol{u}^{k+1} - \sum_i P_i \boldsymbol{\lambda}_1^k -  \boldsymbol{\lambda}_2^k \rm I) \\
\end{aligned}
\end{equation}
4) \textit{The Lagrangian and penalty terms are updated as follows}: \\
\begin{equation}
\begin{aligned}
  {\boldsymbol{\lambda}_1^{k+1}} = \boldsymbol{\lambda}_1^{k} + \boldsymbol{\mu}_1^{k}(\boldsymbol{x}^{k+1}-\Phi \boldsymbol{\alpha}^{k+1})\\
   {\boldsymbol{\lambda}_2^{k+1}} = \boldsymbol{\lambda}_2^{k} + \boldsymbol{\mu}_2^{k}(\boldsymbol{x}^{k+1}-\boldsymbol{u}^{k+1})
\end{aligned}
\end{equation}
\begin{equation}
\begin{aligned}
 \boldsymbol{\mu}_1^{k+1} = \boldsymbol{\rho}_1 \boldsymbol{\mu}_1^{k}\\
 \boldsymbol{\mu}_2^{k+1} = \boldsymbol{\rho}_2 \boldsymbol{\mu}_2^{k}
\end{aligned}
\end{equation} \\
The proposed Low-Rank and Sparsity Plug-and-Play(LRS-PnP) inpainting model is presented in Algorithm \ref{algorithm:LRS-PnP}. 
\begin{algorithm}  
    \caption{(LRS-PnP) Algorithm}
    \label{algorithm:LRS-PnP}
  \begin{algorithmic}[1]
    \REQUIRE masking matrix: $\rm M$, noisy and incomplete HSI: $\boldsymbol{y}$, learned dictionary: $\Phi$. denoiser: $\mathcal{D}$, max iteration: $It_{max}$.
    
    \OUTPUT inpainted HSI image $X$.
    \STATE \textbf{Initialization}: $\boldsymbol{\lambda}_1,\boldsymbol{\lambda}_2,\boldsymbol{\mu}_1,\boldsymbol{\mu}_2, \boldsymbol{\rho}_1,\boldsymbol{\rho}_2$.
      \WHILE{Not Converged}
        \STATE  for $i=1:It_{max}$ do:

               $\boldsymbol{\alpha}^{k+1} = \mathcal{D}(I-\nabla f)(\boldsymbol{\alpha}^{k})$
      
         \STATE ${\boldsymbol{u}^{k+1}} = SVT(\boldsymbol{x}^k+\frac{\boldsymbol{\lambda}_2^k}{\boldsymbol{\mu}_2^k})$
         \STATE update $\boldsymbol{x}$ by \eqref{x_closed_form}.
         \STATE update Lagrangian parameters and penalty terms.
    \ENDWHILE
  \end{algorithmic}
\end{algorithm} \\
Algorithm \ref{algorithm:LRS-PnP-DIP} presents LRS-PnP-DIP, an extension of the LRS algorithm achieved by replacing the SVT with the DIP $f_{\theta}$.
\begin{algorithm}
    \caption{(LRS-PnP-DIP) Algorithm}
    \label{algorithm:LRS-PnP-DIP}
  \begin{algorithmic}[1]
    \REQUIRE masking matrix: $\rm M$, noisy and incomplete HSI: $\boldsymbol{y}$, learned dictionary: $\Phi$. denoiser: $\mathcal{D}$, max iteration: $It_{max}$. DIP: $f_{\theta}$
    \OUTPUT inpainted HSI image $\boldsymbol{x}$.
    \STATE \textbf{Initialization} DIP parameters, $\boldsymbol{\lambda}_1,\boldsymbol{\lambda}_2,\boldsymbol{\mu}_1,\boldsymbol{\mu}_2, \boldsymbol{\rho}_1,\boldsymbol{\rho}_2$.
    \WHILE{Not Converged}
      \STATE  for $i=1:It_{max}$ do:
      
               $\boldsymbol{\alpha}^{k+1} = \mathcal{D}(I-\nabla f)(\boldsymbol{\alpha}^{k})$
      \STATE update $\theta$ in DIP, with the target $\boldsymbol{y}$ and input $\boldsymbol{x}^k + \frac{\boldsymbol{\lambda}_2^k}{\boldsymbol{\mu}_2^k}$.
      \STATE update $\boldsymbol{x}$ by \eqref{x_closed_form}.
      \STATE update Lagrangian parameters and penalty terms.
    \ENDWHILE
  \end{algorithmic}
\end{algorithm}

\section{Convergence Analysis}
In this part, we will show the fixed-point convergence of our proposed LRS-PnP-DIP Algorithm under mild assumptions. Fixed point convergence refers to the type of convergence where the algorithm asymptotically enters into a steady state. To establish our main theorem, we need the following results: \\
\textbf{Definition 1} (Non-Expansive Operator). An operator $T:\Bbb{R}^n \rightarrow \Bbb{R}^n$ is said to be non-expansive if for any $x, y \in \Bbb{R}^n$:
\begin{equation}
\begin{aligned}
     \Vert(T(x)-T(y))\Vert^2 &\le  \Vert(x-y)\Vert^2 
\end{aligned}
\end{equation}
\textbf{Definition 2} ($\theta$-averaged). An operator $T:\Bbb{R}^n \rightarrow \Bbb{R}^n$ is said to be $\theta$-averaged with some $\theta \in (0,1)$ if there exists a non-expansive operator $R$ such that we can write $T = (1-\theta) I+\theta R$\\
\textbf{\setword{Lemma 1}{Lemma 1}}. Let $T:\Bbb{R}^n \rightarrow \Bbb{R}^n$ be $\theta$-averaged for some $\theta \in (0,1)$. Then, for any $x, y \in \Bbb{R}^n$ :
\begin{equation}
\begin{aligned}
     \Vert(T(x)-T(y))\Vert^2 &\le  \Vert(x-y)\Vert^2 \\ &-\frac{1-\theta}{\theta}\Vert((I-T)(x)-(I-T)(y)\Vert^2
\end{aligned}
\end{equation}
Proof of this lemma can be found in [\cite{nair2021fixed}, Lemma 6.1]. \\
\textbf{Definition 3} (Fixed Point). We say that $x^*\in\Bbb{R}^n$ is a fixed point of the operator $T: \Bbb{R}^n \rightarrow \Bbb{R}^n$ if $T(x^*)=x^*$. We denote the set of fixed points by fix($T$). \\
\textbf{Definition 4} ($\beta$-smoothed). Let $f:\Bbb{R}^n \rightarrow \Bbb{R}^n$ be differentiable, we say that $f$ is $\beta$-smooth, if there exists $\beta >0$ such that $\Vert\nabla f(x) - \nabla f(y)\Vert \le \beta \Vert x-y \Vert $ for any $x,y \in \Bbb{R}^n$. \\ 
\textbf{Definition 5} (Strong Convexity). A differentiable function $f$ is said to be strongly convex with modulus $\rho>0$ if $f(x)- \frac{\rho}{2}\Vert x \Vert ^2$ is convex. \\
\textbf{\setword{Lemma 2}{Lemma 2}} (Property of Strong Convexity). Let $f$ be strongly convex with modulus $\rho>0$. Then, for any $x, y \in \Bbb{R}^n$: \\
\begin{equation}
\begin{aligned}
     \left \langle \nabla f(x)-\nabla f(y), x-y\right \rangle &\ge  \rho \Vert(x-y)\Vert^2 
\end{aligned}
\end{equation} 
Proof of the \ref{Lemma 2} can be found in \cite{ekeland1999convex}. \\
We made the following assumptions: \\
\textbf{\setword{Assumption 1}{Assump 1}}: We assume that 1), the Denoiser $\mathcal{D}$ used in the sparse coding step is linear and $\theta$-averaged for some $\theta \in (0, 1)$. 2), the function $f(\boldsymbol{\alpha})=\frac{\boldsymbol{\mu}_1^k}{2} \Vert(\boldsymbol{x}^k + \frac{\boldsymbol{\lambda}_1^k}{\boldsymbol{\mu}_1^k}) - \Phi \boldsymbol{\alpha} \Vert_{2}^2 $ is $\beta$-smoothed. and  3),$ﬁx(\mathcal{D}(I-\nabla f)) \neq \emptyset $.\\
\textit{Remarks}. The $\theta$-averaged property is a weaker assumption compared to the non-expansive Denoiser where most of the existing PnP frameworks have worked with\cite{PnP_Non_Expansive_D,PnP_Non_Expansive_Network}. In fact, the non-expansive assumption is found to be easily violated for denoisers such as BM3D and NLM when used in practice. Nevertheless, the convergence of PnP methods using such denoisers can still be empirically verified. In our experiments, we adopt the modified NLM that satisfies the $\theta$-averaged property by design\cite{PnP_Non_Expansive_D}. \\
\textbf{Lemma 3} (Fixed-Point Convergence of PnP Sparse Coding). If the \ref{Assump 1} holds, Then, for any $\boldsymbol{\alpha}$, and $\rho>\beta/2$. the sequence $\boldsymbol{(\alpha)^k}_{k\ge0}$ generated by step (6) converge to some $\boldsymbol{\alpha}^* \in ﬁx(\mathcal{D}(I-\nabla f)) $
Proof of Lemma 3 can be found in [\cite{nair2021fixed}, Theorem 3.5]. \\
\textbf{\setword{Assumption 2}{Assump 2}}: We assume that the DIP function $f_{\theta}$ is L-Lipschitz bounded: 
\begin{equation}
\begin{aligned}
     \Vert(f_\theta(x)-f_\theta(y))\Vert^2 \le L \Vert(x-y)\Vert^2 
\end{aligned}
\end{equation}
for any $x$,$y$, and $L\le1$. \\
\textit{Remarks}. \ref{Assump 2} guarantees that the trained DIP has Lipschitz constant $L\le1$, this can be achieved by constraining the spectral norm of each layer during training.\\
Now, we are ready to state our main theorem. \\
\textbf{\setword{Theorem 1}{theorem 1}} (Convergence of LRS-PnP-DIP in Lyapunov Sense). If both \ref{Assump 1} and \ref{Assump 2} hold, with penalty $\boldsymbol{\mu}$, and with a L-Lipschitz constrained DIP($L\le1$). Then, there exists a non-increasing function: $H^k = 2\Vert \boldsymbol{x}^k-\boldsymbol{x}^*\Vert^2 +\frac{1}{\boldsymbol{\mu}^2}\Vert \boldsymbol{\lambda}_1^k-\boldsymbol{\lambda}_1^*\Vert^2 +\frac{1}{\boldsymbol{\mu}^2}\Vert \boldsymbol{\lambda}_2^k-\boldsymbol{\lambda}_2^*\Vert^2$, such that all trajectories generated by LRS-PnP-DIP are bounded, and that as $k \rightarrow \infty$, $\Vert \boldsymbol{x}^{k} - \boldsymbol{x}^* \Vert^2 \rightarrow 0$, $\Vert \boldsymbol{\alpha}^{k} -\boldsymbol{\alpha}^* \Vert^2 \rightarrow 0$, and $\Vert \boldsymbol{u}^{k} - \boldsymbol{u}^{*} \Vert^2 \rightarrow 0$. \\.i,e. even if the equilibrium states of $\boldsymbol{x}$, $\boldsymbol{\alpha}$, and $\boldsymbol{u}$ are perturbed, they will finally converge to $\boldsymbol{x}^{*}$, $\boldsymbol{\alpha}^{*}$, and $\boldsymbol{u}^{*}$, respectively. The proposed LRS-PnP-DIP is thus asymptotically stable (The detailed proof of  \ref{theorem 1} is given in Appendix).

\section{Implementation Details}
During our experiments, we evaluated the proposed inpainting model using the Chikusei airborne hyperspectral dataset, which was captured by the Headwall Hyperspec-VNIR-C imaging sensor \cite{Chikusei}. The test HSI image consisted of 128 spectral bands, with each image patch at a particular band having dimensions of 36x36 pixels. We trained the dictionary $\Phi$ with a size of 1296x2000, using only the noisy and incomplete HSI image. This part is a necessary step if we do not have access to the pure spectra. If a standard spectral dataset exists, the learning can be done in the form of a pretraining step. However, we used the input HS image for this task to demonstrate that the proposed algorithms work in a self-supervised setting. Additionally, we introduced Gaussian noise with a fixed $\sigma$ value of 0.12 to the cropped HSI images. As our focus is on inpainting rather than denoising, we maintained a fixed noise strength in all our experiments. We applied the mask $M$ to all spectral bands in the given region, which represented the most challenging case. For the choice of the PnP denoiser, we use both the BM3D and modified Non-local-Mean (NLM) denoisers \cite{PnP_Non_Expansive_D}, the latter enjoys from a non-expansive property that is crucial to the convergence analysis. Our implementation of the Deep Image Prior model followed the same structure as in the Deep Hyperspectral Prior paper \cite{sidorov2019deep}. When training the 1-Lipschitz DIP, we applied Lipschitz regularization to all layers, similar to \cite{henry_gouk_2021regularisation}. In each iteration, we fed the DIP with the latent image from the previous iteration, rather than a random signal. We did not perturb the input to support our convergence analysis, as it is the convention in DIP applications. To avoid manually selecting the total number of iterations in the DIP model, we implemented the early stopping criterion proposed in \cite{wang2021early}, which detects the near-peak PSNR point using windowed moving variance (WMV). The window size and patience number were set to 20 and 100, respectively. The low-rank and sparsity constraints in the equation were weighted with parameters $w_{lr}$ and $w_{s}$, respectively, while the data fidelity term was weighted with $\gamma$. Initially, we set $w_{s}/w_{lr}$ to be 1 and $\gamma$ to be 0.5. It's important to note that the choice of $\gamma$ is highly dependent on the noise level of the observed image. If the noise level is low, the recovered image $X$ should be similar to the noisy observation $Y$, and so the parameter $\gamma$ should be large, and vice versa. The Adam optimizer was used, and the learning rate was set to 0.1. Finally, we used two widely used metrics, namely Mean Signal-to-Noise Ratio (MPSNR) and Mean Structural Similarity (MSSIM), to evaluate the performance of the model in all experiments.

\section{Numerical Results}

\subsection{Replacing SVT with DIP} 
\begin{figure}
\hspace{-6mm}
 \includegraphics[width=.51\textwidth,height=.36\textwidth]{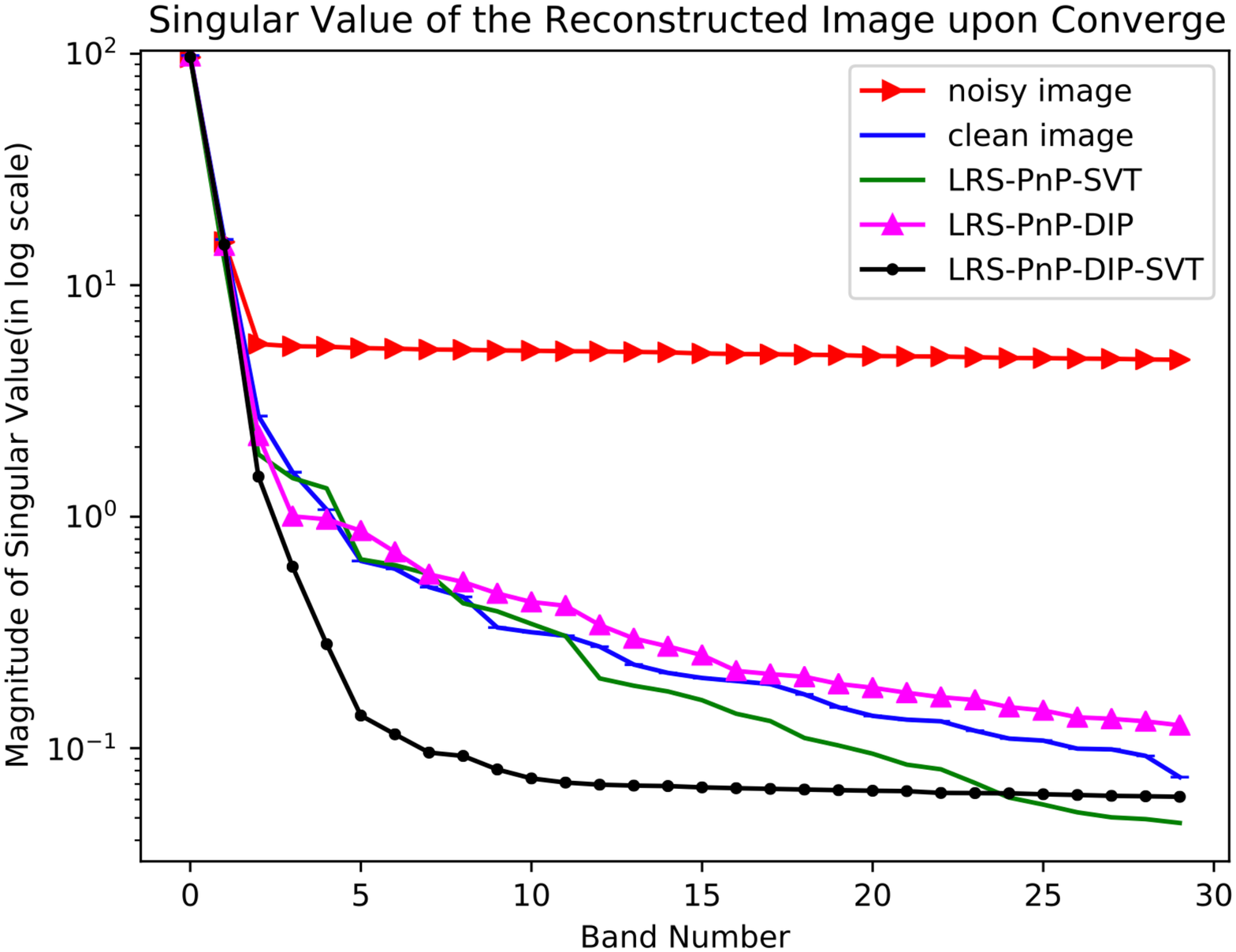}   
  \vspace{-0.4cm}
 \caption{The amplitude of the singular value of the reconstructed image upon converge. The important singular values are captured and preserved by the DIP neural network, even more accurate than the enforced SVT process.}
 \label{fig: singular_value}
\end{figure}

In Algorithm \ref{algorithm:LRS-PnP-DIP}, we propose to solve the low-rank minimisation problem with the DIP. In general, DIP can be applied either alone or combined with the SVT process. However, due to the reason that 1) some small yet potentially important singular values will be removed by the SVT projection, and 2) DIP fits low-frequency components much faster than it fits into the high-frequency noise\cite{wang2021early}, the DIP is only applied prior to the SVT. We compared the singular values of the reconstructed image in Figure \ref{fig: singular_value}, and observe that using a well-trained DIP alone is enough to capture these low-rank details. We thus suggest replacing the whole SVT process with the DIP. To the best of our knowledge, this is the first DIP application on the learning of the low-rank structures in the clean HSIs, Although there is yet no theoretical evidence on how DIP mimic the SVT, we show through experiments that the DIP has the ability to better explore the low-rank subspace, possibly due to its highly non-linear network structure and inductive bias.

\subsection{Convergence} 
\begin{figure}
\hspace{-5mm}
\vspace{-0.1cm}
\includegraphics[width=.53\textwidth,height=.52\textwidth]{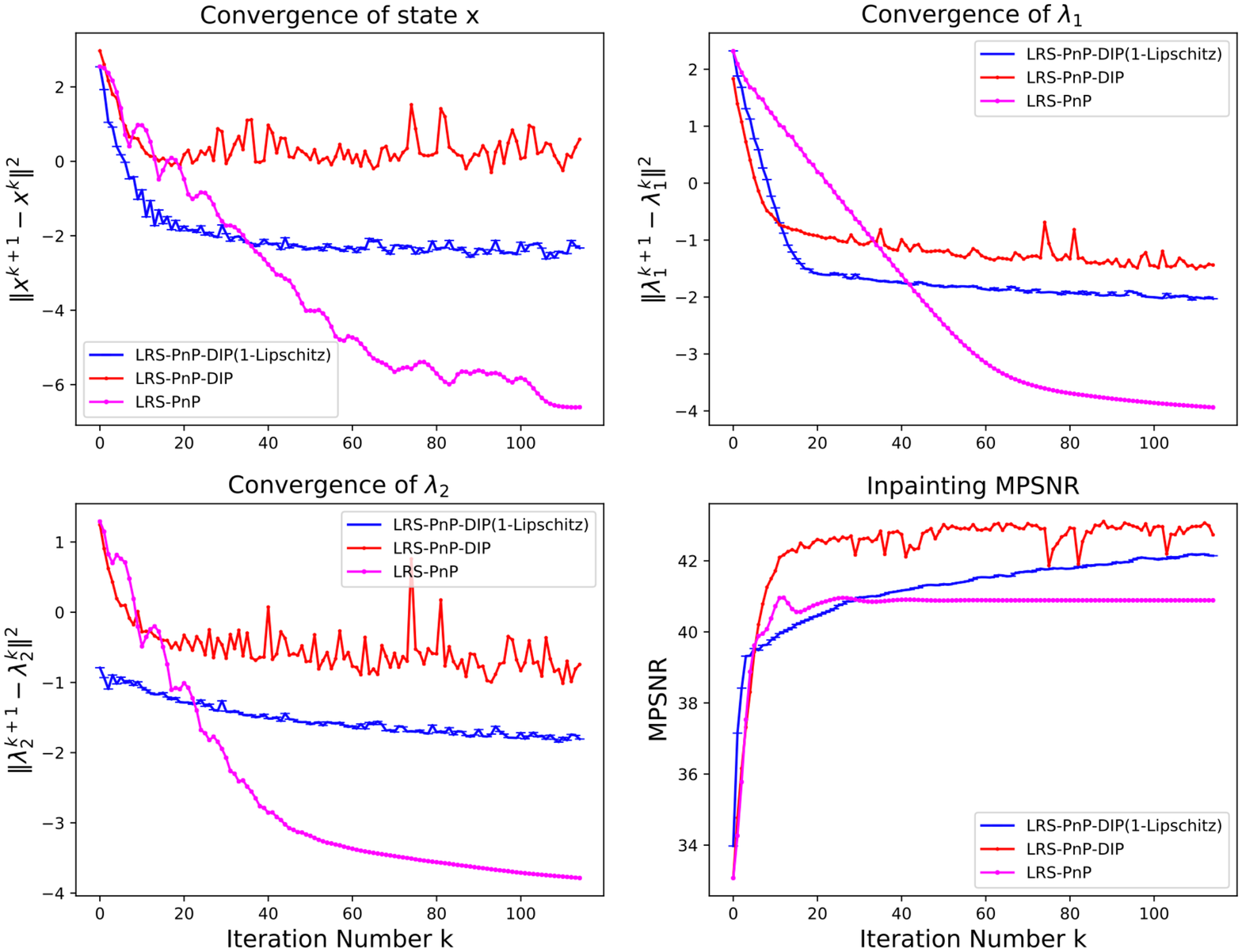} 
 \caption{Empirical converge of LRS-PnP-DIP(1-Lipschitz) with modified NLM denoiser and non expansive/1-Lipschitz DIP. Top Left: successive difference of $\boldsymbol{x}$ in log scale. Top Right: successive difference of $\boldsymbol{\lambda}_1$ in log scale. Bottom Left: successive difference of $\boldsymbol{\lambda}_2$ in log scale. Bottom Right: The inpainting MPSNR vs. Number of Iterations.}
\label{fig: emperical convergence}
\end{figure}

The use of the PnP denoiser and DIP under the HSI settings has been recently investigated in \cite{lai2022deep}. However, the theoretical evidence on the stability of the algorithm is still lacking. To bridge this gap, we verifying the convergence of LRS-PnP-DIP algorithm and its variant LRS-PnP-DIP(1-Lipschitz) in Figure \ref{fig: emperical convergence}. For LRS-PnP-DIP, we use the BM3D denoiser and the DIP without a constraint. For LRS-PnP-DIP(1-Lipschitz), we use modified NLM denoiser and 1-Lipschitz constraint DIP. The latter automatically satisfies \ref{Assump 1} and \ref{Assump 2}, thus the fixed point convergence is guaranteed. Interestingly, we noticed that LRS-PnP-DIP with BM3D and conventional DIP still works pretty well in practice even though there are fluctuations on the state $\boldsymbol{x}$. From the top-right and the bottom-left plots in Figure \ref{fig: emperical convergence}, we deduce that such an instability mainly comes from the process of solving DIP sub-problems rather than the denoising sub-problems, as the primal variable $\boldsymbol{\lambda}_1$ is smoothly converging with BM3D denoisers as well. From the bottom-right plot, we observed that the best results is obtained with LRS-PnP-DIP in terms of reconstruction quality, and the LRS-PnP-DIP(1-Lipschitz) is only slightly lower than its unconstrained counterpart. Our observations on the reduced performance of Lipschitz DIP is in agreement with the existing works \cite{miyato2018spectral,henry_gouk_2021regularisation}, where imposing the Lipschitz constraint during training does hurt the capacity and expressivity of the neural network.  

\subsection{Low-Rank v.s. Sparsity}
\begin{figure}
\hspace{-6mm}
\vspace{-0.6cm}
\includegraphics[width=0.53\textwidth,height=0.4\textwidth,]{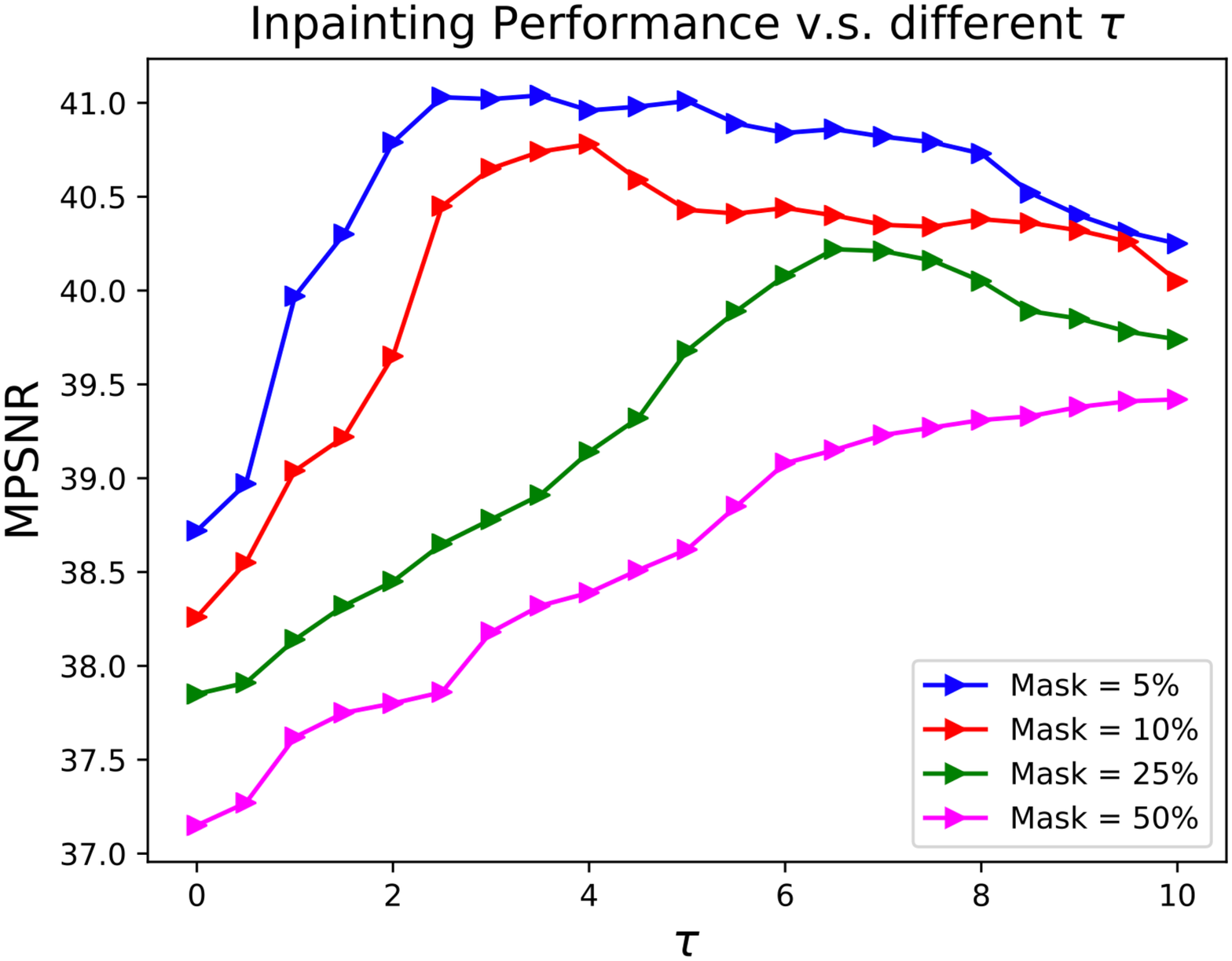} 
\caption{Comparison of MPSNR value of LRS-PnP among different $\tau$($\tau = w_{s}/w_{lr}$) under different masks.}
 \label{fig: sparsity vs low-rank}
\end{figure}

In this section, we will examine the effectiveness of both the sparsity and the low-rank constraint in our proposed LRS-PnP algorithm. Figure \ref{fig: sparsity vs low-rank} shows the inpainting performance of LRS-PnP over different $\tau$, under different masks(in percentage). For example, $\tau=0$ means that the sparsity constraint is disabled, Mask = 50\% means that 50\% of the pixels are totally missing. We can conclude from the results that, 1) the LRS-PnP algorithm works better with both low-rank and sparsity constraints, using either overwhelming sparsity constraint or low-rank constraint alone tends to reduce inpainting performance, and 2) When the percentage of missing pixels increases, more weight on the Sparsity constraint should be put in order to achieve high-quality reconstruction. On one hand, approximation errors may be introduced in the sparse coding step, i,e. step \eqref{alpha} in Algorithm \ref{algorithm:LRS-PnP} and \ref{algorithm:LRS-PnP-DIP}, these errors can potentially be suppressed by the constraint of low-rank. On the other hand, the sparsity constraint exploits the spectral correlations, \textit{i.e.} the performance will be enhanced if spatial information is taken into considerations. 

\begin{figure}
\hspace{-2mm}
\vspace{-0.4cm}
\includegraphics[width=0.50\textwidth,height=0.4\textwidth,]{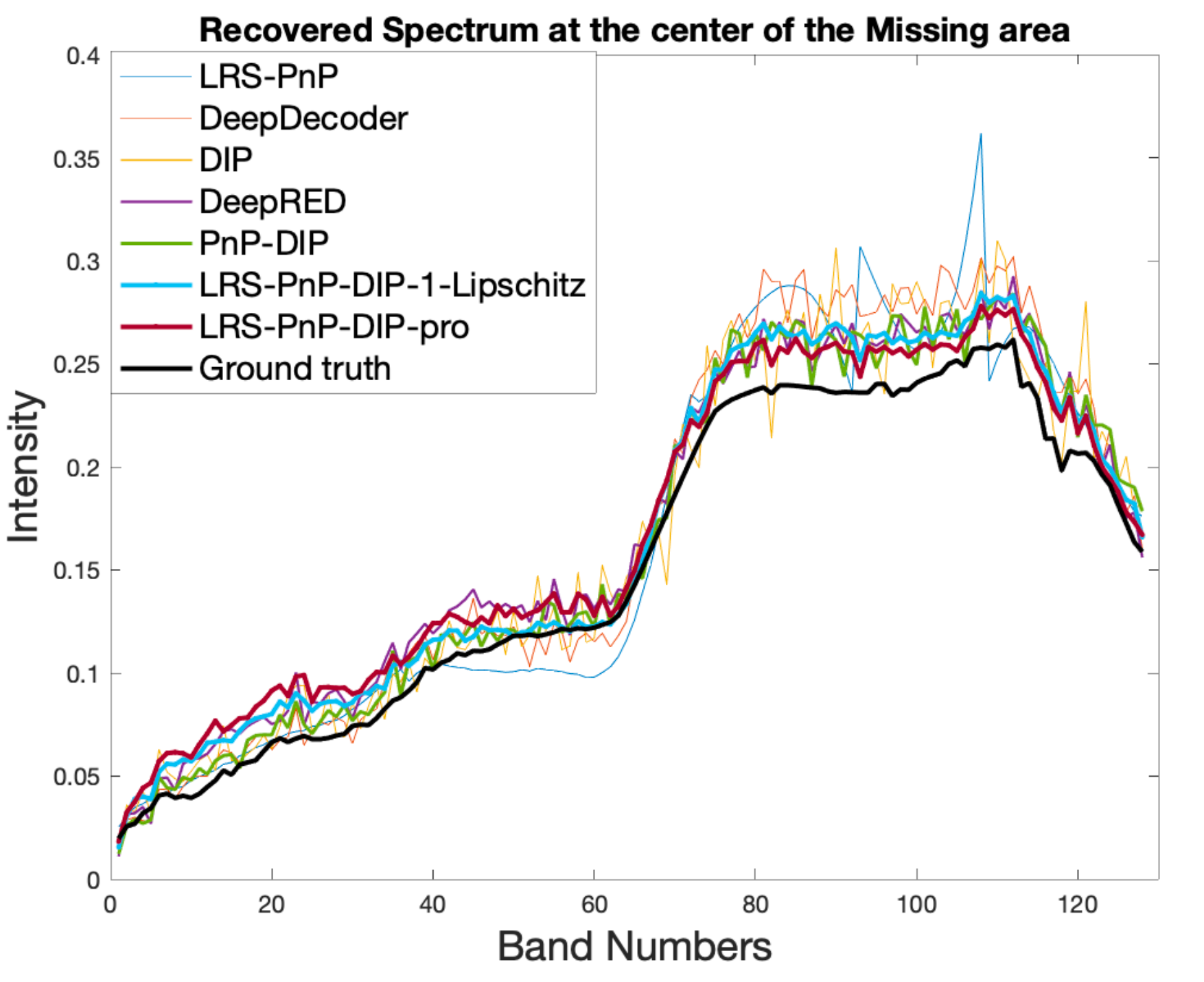} 
\caption{Different algorithms and their recovered spectrum of the center pixel with the assumption that the whole spectrum bands are missing. MPSNR increases from thin to thick line.}
\label{recovered spectrum}
\end{figure}

\begin{table*}
\centering
\scalebox{0.85}{
\begin{tabular}{c rrrrrrrr}
\hline\hline
Methods & Input & LRS-PnP(Ours) & Deep Decoder\cite{Deep_Decoder} & DIP\cite{ulyanov2018deep} & DeepRED\cite{2019_Deep_Red}  &PnP-DIP\cite{pnp_dip} &LRS-PnP-DIP(1-Lip)(Ours) & LRS-PnP-DIP(Ours)\\
\hline 
MPSNR $\uparrow$ & 33.074 & $\boldsymbol{40.9621}$ & 41.0322($\pm$ 0.12)  & 41.3956($\pm$ 0.52) & 41.5765($\pm$ 0.28) & 41.7595($\pm$ 0.25) & $\boldsymbol{41.8023(\pm0.16)}$& $\boldsymbol{42.4880(\pm0.38)}$\\ 
\hline 
MSSIM$\uparrow$ & 0.244 & $\boldsymbol{0.9165}$ & 0.900($\pm 0.01)$ & 0.9102($\pm$ 0.35)   & 0.9121($\pm$ 0.21) & 0.9250($\pm$0.13)  & $\boldsymbol{0.9275(\pm0.12)}$& $\boldsymbol{0.9487(\pm0.20)}$\\ 
\hline

\end{tabular}
}
\label{tab:hresult}

\caption{Comparing the performance with the deep learning methods: the mean and variance over 20 samples are shown here} 
\label{compare DL methods}
\end{table*}

\begin{figure*}   
  \scalebox{1.05}{
  \includegraphics[width=0.95\textwidth,height=0.65\textwidth,]{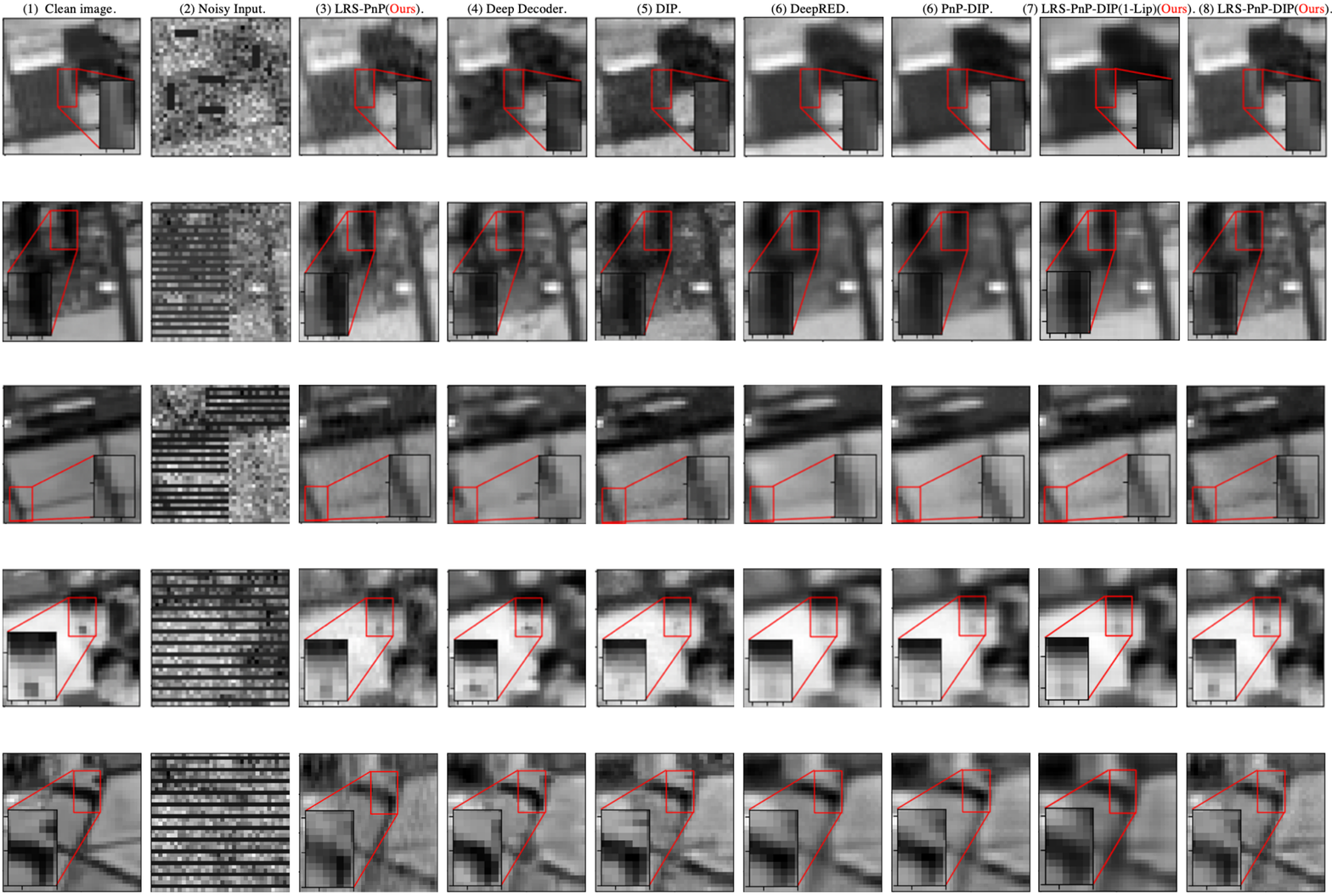}
 }
   \vspace{-1.0cm}
  \caption{Comparison between the proposed algorithms and other learning-based inpainting algorithms. From Left to Right: (1) Clean Image, (2) Input Image, (3) LRS-PnP, (4) Deep Decoder, (5) DIP (6) DeepRED, (7) PnP-DIP, (8) LRS-PnP-DIP(1-Lipschitz) with modified NLM denoiser and 1-Lipschitz Constraint. (8) LRS-PnP-DIP without any constraints. All images are visualized at band 80.}
  \label{reconstruction results}  

\end{figure*}

\begin{table}
\centering
\begin{tabular}{c rrrrr}
\hline\hline
Methods & Input & LRTV\cite{LRTV} & FastHyIn\cite{zhuang2018fast} & LRS-PnP\\
\hline 
MPSNR $\uparrow$ & 33.192 & 39.9878 & 40.5273 & $\boldsymbol{41.0902}$\\ 
\hline 
MSSIM $\uparrow$ & 0.276 & 0.690 & 0.815 & $\boldsymbol{0.9205}$\\ 
\hline
\label{compare_traditional}
\end{tabular}

\caption{Performance Comparing with Traditional Methods in MPSNR and MSSIM metrics} 
\label{compare Traditional methods}
\end{table}

\subsection{Comparison with State-of-the-Art}
We compare the proposed algorithms with the existing traditional methods, including LRTV\cite{LRTV}, FastHyIn\cite{zhuang2018fast}, and learning-based DIP\cite{ulyanov2018deep}, Deep Decoder\cite{Deep_Decoder}, DeepRED\cite{2019_Deep_Red} and PnP-DIP\cite{pnp_dip}. The result is shown in Table \ref{compare DL methods} and Table \ref{compare Traditional methods}. To ensure a fair comparison, we employ U-net as the backbone for DIP, DeepRED, PnP-DIP, and proposed LRS-PnP-DIP/LRS-PnP-DIP(1-Lipschitz), and aggregate the results over 20 experiments. We adjust the network of Deep Decoder to have a 3-layer decoder to fit the size of our HSI image while keeping the remaining structure the same as proposed in the original paper \cite{Deep_Decoder}. The best result is obtained through fine-tuning the learning rates and the noise standard deviations. For the comparison with FastHyIn and LRTV, one-fourth of the pixels are randomly masked using a uniform distribution for the missing bands. The recovered spectrum of the center of the missing area and the reconstructed image are shown in Figure \ref{recovered spectrum} and Figure \ref{reconstruction results}, respectively.
From Figure \ref{reconstruction results}, it is evident that LRS-PnP produces a more consistent and realistic spectrum in the missing region, compared to the learning-based methods, despite having the lowest overall MPSNR. Note the LRS-PnP algorithm is specifically designed to recover the entire missing band, which corresponds to the most challenging scenario in practice, known as a dead pixel. Traditional algorithms such as LRTV and FastHyIn are not able to handle this scenario, whereas our algorithm can. We also observed that the LRS-PnP algorithm has the ability to capture local structures such as abrupt material changes, as evident in the first and last two images of Figure \ref{reconstruction results}. In contrast, all other learning-based methods such as DIP, Deep Decoder, DeepRED, and PnP-DIP produce smoother reconstruction results in the non-missing regions compared to LRS-PnP. However, there are noticeable distortions and artifacts in the missing regions in the results of those methods, possibly due to the reason that DIP generally focuses on learning global features and characteristics. In practice, DIP is often used with additional regularisers, \textit{e.g.} Total Variation(TV), to preserve more details\cite{2019_DIP_TV}. Despite this fact, the LRS-PnP algorithm remains highly competitive when compared to other learning-based methods. We then combine the LRS-PnP algorithm with the DIP to further improve the performance. The resulting LRS-PnP-DIP and its variant LRS-PnP-DIP(1-Lipschitz) are shown in the last two columns in Figure \ref{reconstruction results}. It is observed that even if LRS-PnP-DIP(1-Lipschitz) is highly restricted due to the layer-wise spectral normalisation, its inpainting performance is on par with the PnP-DIP. The inpainting result of LRS-PnP-DIP algorithm is both visually and qualitatively better than those of other state-of-the-art inpainting algorithms. We also noticed that LRS-PnP-DIP works better than its SVT counterparts. We thus suspect that DIP could learn better from the noisy images and from the small singular values which would otherwise be discarded by SVT.

\section{Conclusion}
The LRS-PnP algorithm, along with its extension self-supervised LRS-PnP-DIP, are innovative hyperspectral inpainting techniques capable of handling missing pixels in the noisy and incomplete HS images, even in the most challenging scenarios where entire spectral bands are absent. The new methods exploit spectral and spatial redundancy of HSIs and require no training data except, the test image. A comparison of LRS-PnP and LRS-PnP-DIP with the state-of-the-art algorithms on a real dataset indicates that LRS-PnP provides comparable performance, while LRS-PnP-DIP outperforms other learning-based methods. Unlike previous works where the convergence of the algorithm is either under-studies or only empirically verified \cite{2021_DIP_In_Loop,pnp_dip,DIP-TV,lai2022deep,wu2022adaptive}, the theoretical evidences are provided which demonstrate a convergence guarantee under some mild assumptions. Since the performance of DIP is highly sensitive to the network structures \cite{ulyanov2018deep}, in the future work we would like to explore a broader variety of generative models investigate their potentials in solving HSI inpainting problems.

\bibliographystyle{unsrt}

\bibliography{Bibliography} 

\section{Appendix} 
{\centering\section*{Proof of Theorem 1}}
The proof relies on the Lyapunov stability theory which is the heart of the dynamic system analysis\cite{shevitz1994lyapunov}. The Lyapunov stability theory can be categorized into a)the indirect method, which analyzes the convergence through the system state equation, and b)the direct method, which explicitly describes the behaviour of the system's trajectories and its convergence by making use of the Lyapunov function. We refer the readers to \cite{bof2018lyapunov} for more detailed definition of the Lyapunov function(specifically, Theorem 1.2 and Theorem 3.3). 
In some contexts, it is also known as the energy or dissipative function\cite{hill1976stability}. Compared to the former, the direct method is more appealing as the convergence can be established by only showing the existence of such a function. In this proof, we will start by defining a function for our proposed LRS-PnP-DIP algorithm, and prove its validity. \\
Let $H^k = 2\Vert \boldsymbol{x}^k-\boldsymbol{x}^*\Vert^2 +\frac{1} {\boldsymbol{\mu}^2}\Vert \boldsymbol{\lambda}_1^k-\boldsymbol{\lambda}_1^*\Vert^2 +\frac{1} {\boldsymbol{\mu}^2}\Vert \boldsymbol{\lambda}_2^k-\boldsymbol{\lambda}_2^*\Vert^2$, for a non-zero $\boldsymbol{\mu}$. \\
\textit{Remarks}. This design follows the similar structures as in the original convergence proof of the ADMM algorithm\cite{boyd2011distributed} and in a recent work\cite{zhang2019fundamental_ADMM}. Here, $H^k$ is a function of the system's state change, which is by design, non-negative. The first two assumptions in Theorem 1.2\cite{bof2018lyapunov} automatically hold. Thus, we only need to show that the proposed candidate $H^k$ is a non-increasing function in order to be a valid Lyapunov function.
More specifically, we will show that $H^k$ is a
decreasing function which satisfies: \\

\begin{equation}
\begin{aligned}
  H^k-H^{k+1} \ge C
\end{aligned}
\end{equation}
where C is a positive constant. \\
Firstly, recall that the LRS-PnP-DIP algorithm takes the following update steps:
\begin{equation}  \label{alpha_k+1}
\begin{aligned}
 {\boldsymbol{\alpha}^{k+1}}=\mathcal{T}(\boldsymbol{x}^{k} + \frac{\boldsymbol{\lambda}_1^k}{\boldsymbol{\mu}})
\end{aligned}
\end{equation}

\begin{equation}
\begin{aligned}
  {\boldsymbol{u}^{k+1}}=f_\theta(\boldsymbol{x}^{k}+\frac{\boldsymbol{\lambda}_2^k}{\boldsymbol{\mu}})
\end{aligned}
\end{equation}
\begin{equation} \label{x_k+1}
\begin{aligned}
  {\boldsymbol{x}^{k+1}} = \argmin_{\boldsymbol{x}} \Vert \boldsymbol{y} -\rm M\boldsymbol{x} \Vert_{2}^2 + \frac{\boldsymbol{\mu}}{2}\Vert(\boldsymbol{x} + \frac{\boldsymbol{\lambda}_1^k}{\boldsymbol{\mu}})-\Phi\boldsymbol{\alpha}^{k+1} \Vert_{2}^2 \\
  + \frac{\boldsymbol{\mu}}{2}\Vert (\boldsymbol{x}+\frac{\boldsymbol{\lambda}_2^k}{\boldsymbol{\mu}}) - \boldsymbol{u}^{k+1} \Vert_{2}^2 \\
\end{aligned}
\end{equation}
\begin{equation} \label{lagrangian}
\begin{aligned}
  {\boldsymbol{\lambda}_1^{k+1}} = \boldsymbol{\lambda}_1^{k} + \boldsymbol{\mu}(\boldsymbol{x}^{k+1}-\Phi \boldsymbol{\alpha}^{k+1})\\
   {\boldsymbol{\lambda}_2^{k+1}} = \boldsymbol{\lambda}_2^{k} + \boldsymbol{\mu}(\boldsymbol{x}^{k+1}-\boldsymbol{u}^{k+1})
\end{aligned}
\end{equation}\\
We define $\boldsymbol{x}_e^{k}=\boldsymbol{x}^{k}-\boldsymbol{x}^{*}$, $\boldsymbol{u}_e^{k}=\boldsymbol{u}^{k}-\boldsymbol{u}^{*}$, $\boldsymbol{\alpha_e}^{k}=\boldsymbol{\alpha}^{k}-\boldsymbol{\alpha}^{*}$, $\boldsymbol{\lambda}_{1e}^{k}=\boldsymbol{\lambda}_1^{k}-\boldsymbol{\lambda}_1^{*}$, $\boldsymbol{\lambda}_{2e}^{k}=\boldsymbol{\lambda}_2^{k}-\boldsymbol{\lambda}_2^{*}$, in the subsequent proof for simplicity. This results in: \\
\begin{equation} \label{define_for_simplicity}
\left\{
\begin{aligned}
\boldsymbol{x}_e^{k+1}=\boldsymbol{x}^{k+1}-\boldsymbol{x}^{*} \\
\boldsymbol{u}_e^{k+1}=\boldsymbol{u}^{k+1}-\boldsymbol{u}^{*} \\
\boldsymbol{\alpha}_e^{k+1}=\boldsymbol{\alpha}^{k+1}-\boldsymbol{\alpha}^{*} \\
\boldsymbol{\lambda}_{1e}^{k+1}=\boldsymbol{\lambda}_1^{k+1}-\boldsymbol{\lambda}_1^{*} \\
\boldsymbol{\lambda}_{2e}^{k+1}=\boldsymbol{\lambda}_2^{k+1}-\boldsymbol{\lambda}_2^{*}
\end{aligned}
\right.
\end{equation}\\
We take the equation \eqref{x_k+1} as our starting point and denote the first term $ \Vert \boldsymbol{y} -\rm M\boldsymbol{x} \Vert_{2}^2$ as $f(\boldsymbol{x})$. The first-order optimality of equation \eqref{x_k+1} implies: \\
\begin{equation}
\begin{aligned} \label{1_order_op}
 \nabla f\boldsymbol{(x)} &+\boldsymbol{\mu}(\boldsymbol{x}+\frac{\boldsymbol{\lambda}_1^k}{\boldsymbol{\mu}}-\Phi \boldsymbol{\alpha}^{k+1}) \\
 &+\boldsymbol{\mu}(\boldsymbol{x}+\frac{\boldsymbol{\lambda}_2^k}{\boldsymbol{\mu}}-\boldsymbol{u}^{k+1}) =0
\end{aligned}
\end{equation}
Since the minimizer $\boldsymbol{x}^{k+1}$ satisfies \eqref{1_order_op}, we plug it in equation \eqref{1_order_op} to get: \\
\begin{equation}
\begin{aligned} \label{delta_x_k+1}
 \nabla f\boldsymbol{(x)}^{k+1} +\boldsymbol{\mu}(\boldsymbol{x}^{k+1}-\Phi \boldsymbol{\alpha}^{k+1}+\frac{\boldsymbol{\lambda}_1^k}{\boldsymbol{\mu}}) \\
 +\boldsymbol{\mu}(\boldsymbol{x}^{k+1}-\boldsymbol{u}^{k+1}+\frac{\boldsymbol{\lambda}_2^k}{\boldsymbol{\mu}})  \\
 \overset{\eqref{lagrangian}}{=}\nabla f\boldsymbol{(x)}^{k+1} +\boldsymbol{\mu}(\frac{\boldsymbol{\lambda}_1^{k+1}-\boldsymbol{\lambda}_1^k}{\boldsymbol{\mu}}+\frac{\boldsymbol{\lambda}_1^k}{\boldsymbol{\mu}}) \\
 +\boldsymbol{\mu}(\frac{\boldsymbol{\lambda}_2^{k+1}-\boldsymbol{\lambda}_1^k}{\boldsymbol{\mu}}+\frac{\boldsymbol{\lambda}_2^k}{\boldsymbol{\mu}}) \\
 =\nabla f\boldsymbol{(x)}^{k+1}+\boldsymbol{\lambda}_1^{k+1}+\boldsymbol{\lambda}_2^{k+1} =0
\end{aligned}
\end{equation}
the critical point satisfies, i,e $k\rightarrow \infty$: \\
\begin{equation}
\begin{aligned}\label{delta_x_*}
\nabla f\boldsymbol{(x)}^{*}+\boldsymbol{\lambda}_1^{*}+\boldsymbol{\lambda}_2^{*} =0
\end{aligned}
\end{equation}
Due to the strong convexity of $f(\boldsymbol{x})$, using \ref{Lemma 2} with $\boldsymbol{x}=\boldsymbol{x}^{k+1}$ and $\boldsymbol{y}=\boldsymbol{x}^{*}$ yields: \\
\begin{equation}
\begin{aligned}
   \left \langle \nabla f\boldsymbol{(x)}^{k+1}-\nabla f\boldsymbol{(x)}^{*}, \boldsymbol{x}^{k+1}-\boldsymbol{x}^{*}\right \rangle  &\ge  \rho \Vert(\boldsymbol{x}^{k+1}-\boldsymbol{x}^{*})\Vert^2 
\end{aligned}
\end{equation}
Combining the above inequality with equation \eqref{delta_x_k+1} and \eqref{delta_x_*}, we have: \\
\begin{equation} \label{lambda_inequality}
\begin{aligned}
   \left \langle -\boldsymbol{\lambda}_{1e}^{k+1}-\boldsymbol{\lambda}_{2e}^{k+1}, \boldsymbol{x}_e^{k+1} \right \rangle &\ge  \rho \Vert\boldsymbol{x}_e^{k+1}\Vert^2 
\end{aligned}
\end{equation}
Secondly, using the \ref{Assump 2} that the DIP $f_\theta(z)$ is L-Lipschitz with $L\le1$, and $\boldsymbol{x}=\boldsymbol{x}^{k}+\frac{\boldsymbol{\lambda}_2^k}{\boldsymbol{\mu}}$, $\boldsymbol{y}=\boldsymbol{x}^{*}+\frac{\boldsymbol{\lambda}_2^*}{\boldsymbol{\mu}}$, we get:\\
\begin{equation}\label{u_inequality}
\begin{aligned}
     \Vert\boldsymbol{u}^{k+1}-\boldsymbol{u}^{*}\Vert^2 \le  \Vert(\boldsymbol{x}^{k}-\boldsymbol{x}^{*})+ \frac{\boldsymbol{\lambda}_2^{k}-\boldsymbol{\lambda}_2^*}{\boldsymbol{\mu}}\Vert^2 \\
     or \Vert \boldsymbol{u}_{e}^{k+1}\Vert^2 \le \Vert \boldsymbol{x}_e^{k}+ \frac{\boldsymbol{\lambda}_{2e}^{k}}{\boldsymbol{\mu}}\Vert^2 \\
\end{aligned}
\end{equation}
Thirdly, using the  \ref{Assump 1} and the resulting \ref{Lemma 1} that the operator $\mathcal{T}$ used in the $\boldsymbol{\alpha}$ updating step \eqref{alpha_k+1} is $\theta$-averaged, and  $\boldsymbol{x}=\boldsymbol{x}^{k}+\frac{\boldsymbol{\lambda}_1^k}{\boldsymbol{\mu}}$, $\boldsymbol{y}=\boldsymbol{x}^{*}+\frac{\boldsymbol{\lambda}_1^*}{\boldsymbol{\mu}}$, we have:
\begin{equation}\label{alpha_inequality}
\begin{aligned}
     \Vert\boldsymbol{\alpha}^{k+1}-\boldsymbol{\alpha}^{*}\Vert^2 \le  \Vert(\boldsymbol{x}^{k}-\boldsymbol{x}^{*})+ \frac{\boldsymbol{\lambda}_1^{k}-\boldsymbol{\lambda}_1^*}{\boldsymbol{\mu}}\Vert^2 \\
     or \Vert \boldsymbol{\alpha}_{e}^{k+1}\Vert^2 \le \Vert \boldsymbol{x}_e^{k}+ \frac{\boldsymbol{\lambda}_{1e}^{k}}{\boldsymbol{\mu}}\Vert^2 \\
\end{aligned}
\end{equation}
Now, we multiply \eqref{lambda_inequality} both side by $\frac{2}{\boldsymbol{\mu}}$, and gather the resulting inequality with \eqref{u_inequality} and \eqref{alpha_inequality}:
\begin{equation}\label{gather_three}
\left\{
\begin{aligned}
   -\frac{2}{\boldsymbol{\mu}}\left \langle \boldsymbol{\lambda}_{1e}^{k+1}, \boldsymbol{x}_e^{k+1} \right \rangle - \frac{2}{\boldsymbol{\mu}}\left \langle \boldsymbol{\lambda}_{2e}^{k+1}, \boldsymbol{x}_e^{k+1} \right \rangle &\ge  \frac{2}{\boldsymbol{\mu}}\rho \Vert\boldsymbol{x}_e^{k+1}\Vert^2 \\
   \Vert \boldsymbol{x}_e^{k}+ \frac{\boldsymbol{\lambda}_{2e}^{k}}{\boldsymbol{\mu}}\Vert^2 \ge  \Vert \boldsymbol{u}_{e}^{k+1} \Vert^2 \\
   \Vert \boldsymbol{x}_e^{k}+ \frac{\boldsymbol{\lambda}_{1e}^{k}}{\boldsymbol{\mu}}\Vert^2 \ge \Vert \boldsymbol{\alpha}_{e}^{k+1} \Vert^2 
\end{aligned}
\right.
\end{equation}
We put the $\boldsymbol{\mu}$ inside the left-hand-side of the first inequality, and add them all gives: 
\begin{equation}
\begin{aligned}
   -2\left \langle \frac{\boldsymbol{\lambda}_{1e}^{k+1}}{{\boldsymbol{\mu}}}, \boldsymbol{x}_e^{k+1} \right \rangle - 2\left \langle \frac{\boldsymbol{\lambda}_{2e}^{k+1}}{{\boldsymbol{\mu}}}, \boldsymbol{x}_e^{k+1} \right \rangle  +
   \Vert \boldsymbol{x}_e^{k}+ \frac{\boldsymbol{\lambda}_{2e}^{k}}{\boldsymbol{\mu}}\Vert^2 \\
   + \Vert \boldsymbol{x}_e^{k} + \frac{\boldsymbol{\lambda}_{1e}^{k}}{\boldsymbol{\mu}}\Vert^2 \ge \frac{2}{\boldsymbol{\mu}}\rho \Vert\boldsymbol{x}_e^{k+1}\Vert^2+\Vert \boldsymbol{u}_{e}^{k+1} \Vert^2+\Vert \boldsymbol{\alpha}_{e}^{k+1} \Vert^2 
\end{aligned}
\end{equation}
which can be written as:
\begin{equation}
\begin{aligned}
   -(\frac{1}{\boldsymbol{\mu}^2}\Vert\boldsymbol{\lambda}_{1e}^{k+1}\Vert^2+ \Vert \boldsymbol{x}_{e}^{k+1}\Vert^2- \Vert \frac{\boldsymbol{\lambda}_{1e}^{k+1}}{{\boldsymbol{\mu}}}-\boldsymbol{x}_{e}^{k+1}\Vert^2) \\
   -(\frac{1}{\boldsymbol{\mu}^2}\Vert\boldsymbol{\lambda}_{2e}^{k+1}\Vert^2+ \Vert \boldsymbol{x}_{e}^{k+1}\Vert^2- \Vert \frac{\boldsymbol{\lambda}_{2e}^{k+1}}{{\boldsymbol{\mu}}}-\boldsymbol{x}_{e}^{k+1}\Vert^2) \\
   +\frac{1}{\boldsymbol{\mu}^2}\Vert\boldsymbol{\lambda}_{2e}^{k}\Vert^2+ \Vert \boldsymbol{x}_{e}^{k}\Vert^2+ 2\left \langle \frac{\boldsymbol{\lambda}_{2e}^{k}}{\boldsymbol{\mu}},\boldsymbol{x}_{e}^{k}\right \rangle\\
   +\frac{1}{\boldsymbol{\mu}^2}\Vert\boldsymbol{\lambda}_{1e}^{k}\Vert^2+ \Vert \boldsymbol{x}_{e}^{k}\Vert^2+ 
   2\left \langle \frac{\boldsymbol{\lambda}_{1e}^{k}}{\boldsymbol{\mu}},\boldsymbol{x}_{e}^{k}\right \rangle \\
   \ge \frac{2}{\boldsymbol{\mu}}\rho \Vert\boldsymbol{x}_e^{k+1}\Vert^2+\Vert \boldsymbol{u}_{e}^{k+1} \Vert^2+\Vert \boldsymbol{\alpha}_{e}^{k+1} \Vert^2 
\end{aligned}
\end{equation}
After re-arrangement, we get:
\begin{equation} \label{recover_H_k}
\begin{aligned}
   2(\Vert \boldsymbol{x}_{e}^{k}\Vert^2 - \Vert \boldsymbol{x}_{e}^{k+1}&\Vert^2)+
   \frac{1}{\boldsymbol{\mu}^2}(\Vert \boldsymbol{\lambda}_{1e}^{k}\Vert^2-\Vert \boldsymbol{\lambda}_{1e}^{k+1}\Vert^2) 
   +\frac{1}{\boldsymbol{\mu}^2}(\Vert \boldsymbol{\lambda}_{2e}^{k}\Vert^2 \\
   &-\Vert \boldsymbol{\lambda}_{2e}^{k+1}\Vert^2) \\
   &\ge \frac{2}{\boldsymbol{\mu}}\rho \Vert\boldsymbol{x}_e^{k+1}\Vert^2+\Vert \boldsymbol{u}_{e}^{k+1} \Vert^2+\Vert \boldsymbol{\alpha}_{e}^{k+1} \Vert^2 \\
   &+\Vert \frac{\boldsymbol{\lambda}_{1e}^{k+1}}{\boldsymbol{\mu}}-\boldsymbol{x}_{e}^{k+1}\Vert^2+
   \Vert \frac{\boldsymbol{\lambda}_{2e}^{k+1}}{\boldsymbol{\mu}}-\boldsymbol{x}_{e}^{k+1}\Vert^2\\
   &-2\left \langle \frac{\boldsymbol{\lambda}_{2e}^{k}}{\boldsymbol{\mu}},\boldsymbol{x}_{e}^{k}\right \rangle -
2\left \langle \frac{\boldsymbol{\lambda}_{1e}^{k}}{\boldsymbol{\mu}},\boldsymbol{x}_{e}^{k}\right \rangle \\
\end{aligned}
\end{equation}
Recall that:\\
\begin{equation} \label{recall_H_k}
\begin{aligned}
H^k& - H^{k+1} \\
&= 2\Vert \boldsymbol{x}^k-\boldsymbol{x}^*\Vert^2 +\frac{1} {\boldsymbol{\mu}^2}\Vert \boldsymbol{\lambda}_1^k-\boldsymbol{\lambda}_1^*\Vert^2 +\frac{1} {\boldsymbol{\mu}^2}\Vert \boldsymbol{\lambda}_2^k -
\boldsymbol{\lambda}_2^*\Vert^2& \\
&-2\Vert \boldsymbol{x}^{k+1}-\boldsymbol{x}^*\Vert^2 -\frac{1} {\boldsymbol{\mu}^2}\Vert \boldsymbol{\lambda}_1^{k+1}-\boldsymbol{\lambda}_1^*\Vert^2 -\frac{1} {\boldsymbol{\mu}^2}\Vert \boldsymbol{\lambda}_2^{k+1}-\boldsymbol{\lambda}_2^*\Vert^2 \\
&\overset{\eqref{define_for_simplicity}}{=}2(\Vert \boldsymbol{x}_{e}^{k}\Vert^2 - \Vert \boldsymbol{x}_{e}^{k+1}\Vert^2) +
   \frac{1}{\boldsymbol{\mu}^2}(\Vert \boldsymbol{\lambda}_{1e}^{k}\Vert^2-\Vert \boldsymbol{\lambda}_{1e}^{k+1}\Vert^2) \\
   &+\frac{1}{\boldsymbol{\mu}^2}(\Vert \boldsymbol{\lambda}_{2e}^{k}\Vert^2-\Vert \boldsymbol{\lambda}_{2e}^{k+1}\Vert^2)
\end{aligned} \\
\end{equation}
It can be seen that the left-hand-side of inequality \eqref{recover_H_k} recovers exactly $H^k-H^{k+1}$. To make $H^k$ a non-increasing function, we require the entire right-hand-side to be non-negative, it is thus sufficient to show that the last two terms $-2\left \langle \frac{\boldsymbol{\lambda}_{2e}^{k}}{\boldsymbol{\mu}},\boldsymbol{x}_{e}^{k}\right \rangle -
2\left \langle \frac{\boldsymbol{\lambda}_{1e}^{k}}{\boldsymbol{\mu}},\boldsymbol{x}_{e}^{k}\right \rangle$ is non-negative. This is straightforward to show if we plug in $k=k-1$ into the first line of \eqref{gather_three}, to get:
\begin{equation}
\begin{aligned}
   H^k-H^{k+1}
   &\ge \frac{2}{\boldsymbol{\mu}}\rho \Vert\boldsymbol{x}_e^{k+1}\Vert^2+\Vert \boldsymbol{u}_{e}^{k+1} \Vert^2+\Vert \boldsymbol{\alpha}_{e}^{k+1} \Vert^2 \\
   &+\Vert \frac{\boldsymbol{\lambda}_{1e}^{k+1}}{\boldsymbol{\mu}}-\boldsymbol{x}_{e}^{k+1}\Vert^2+ 
   \Vert \frac{\boldsymbol{\lambda}_{2e}^{k+1}}{\boldsymbol{\mu}}-\boldsymbol{x}_{e}^{k+1}\Vert^2\\
   &+\frac{2}{\boldsymbol{\mu}}\rho \Vert\boldsymbol{x}_e^{k}\Vert^2\\
   &\ge 0
\end{aligned}
\end{equation}
Now, if we add both side of \eqref{recall_H_k}, from $k=0$ to $k=\infty$, it follows that:
\begin{equation}
\begin{aligned}
  \sum\limits_{k=0}^{\infty} \frac{2}{\boldsymbol{\mu}}\rho \Vert\boldsymbol{x}_e^{k+1}\Vert^2+ \sum\limits_{k=0}^{\infty}\Vert \boldsymbol{u}_{e}^{k+1} \Vert^2 &+ \sum\limits_{k=0}^{\infty}\Vert \boldsymbol{\alpha}_{e}^{k+1} \Vert^2  \\
  &\le H^0-H^{\infty} < \infty
\end{aligned}
\end{equation}
We can conclude that the sequence ${\boldsymbol{x}_e^{k+1}}$, {$\boldsymbol{u}_e^{k+1}$} and {$\boldsymbol{\alpha}_e^{k+1}$} are all bounded sequences due to Lyapunov theorem. That is, as $k\rightarrow \infty$: \\
\begin{equation}
\begin{aligned}
  \lim_{k\to\infty} \Vert\boldsymbol{x}^{k+1}-\boldsymbol{x}^{*}\Vert^2 \rightarrow 0 \\
  \lim_{k\to\infty} \Vert\boldsymbol{u}^{k+1}-\boldsymbol{u}^{*}\Vert^2 \rightarrow 0 \\
  \lim_{k\to\infty} \Vert\boldsymbol{\alpha}^{k+1}-\boldsymbol{\alpha}^{*}\Vert^2 \rightarrow 0 
\end{aligned}
\end{equation}
Thus the iterates generated by LRS-PnP-DIP algorithm converge to the critical point ($\boldsymbol{x}^{*},\boldsymbol{u}^{*},\boldsymbol{\alpha}^{*}$) with a sufficiently large k, and all the trajectories all bounded. The LRS-PnP-DIP is also asymptotically stable by Theorem 3.3 \cite{bof2018lyapunov}.

\end{document}